\documentclass[12pt,a4paper]{JHEP}
\usepackage{epsfig}
\usepackage{array,cite,amsmath,amssymb}
\newcommand{\newc}{\newcommand}
\newc{\inmath}[1] {\ifmmode#1\else$#1$\fi}
\newc{\definmath}[2] {\def#1{\ifmmode#2\else$#2$\fi}}
\newc{\gev}{\,GeV}
\newc{\mev}{\,MeV}
\newc{\ra}{\rightarrow}
\definmath{\rpv}{\mathrm{\not\!R_p}}
\definmath{\rp}{\mathrm{R_p}}
\newc{\real}{\mathcal{R}e}
\newc{\alsm}{{\displaystyle \sum_{\alpha=1,2}}}
\newc{\besm}{{\displaystyle \sum_{\beta=1,2}}}
\newc{\al}{\alpha}
\definmath{\lampp}{\lambda^{\prime \prime}}
\newc{\sgn}{\mr{sgn}\,}
\newc{\be}{\beta}
\newc{\ga}{\gamma}
\newc{\de}{\delta}
\newc{\sla}{\!\!\!\!\!\not\:\:\!}
\newc{\slab}{\!\!\!\!\!\not\,\,\,}
\newc{\slac}{\!\!\!\!\!\!\!\not\,\,\,\,}
\newc{\met}{$\not\!\!E_T$}
\newc{\cw}{\cos\theta_W}
\newc{\sw}{\sin\theta_W}
\newc{\ssw}{\sin^2\theta_W}
\newc{\ccw}{\cos^2\theta_W}
\newc{\cbe}{\cos\beta}
\newc{\sbe}{\sin\beta}
\newc{\ort}{\frac1{\sqrt{2}}}
\newc{\sh}{\hat{s}}
\newc{\uh}{\hat{u}}
\newc{\tha}{\hat{t}}
\newc{\sa}{\sin\al}
\newc{\ca}{\cos\al}
\newc{\mz}{M_{\mr{Z}}}
\newc{\mw}{M_{\mr{W}}}
\definmath\gsim{\,\,\rlap{\raise 3pt\hbox{$>$}}{\lower 3pt\hbox{$\sim$}}\,\,}
\definmath\lsim{\,\,\rlap{\raise 3pt\hbox{$<$}}{\lower 3pt\hbox{$\sim$}}\,\,}
\definmath{\bv}{\mathrm{\not\!B}}
\definmath{\lv}{\mathrm{\not\!L}}
\newc{\beq}{\begin{equation}}
\newc{\eeq}{\end{equation}}
\newc{\ie}{{\it i.e.\/}\ }
\definmath{\lam}{\lambda}
\definmath{\cht}{\tilde{\chi}}
\definmath{\chgone}{\cht^+_1}
\definmath{\ntlone}{\cht^0_1}
\definmath{\ntltwo}{\cht^0_2}
\definmath{\sslr} {\tilde{l}_{R}}
\definmath{\glt}{\tilde{\rm{g}}}
\definmath{\upt}{\tilde{\rm{u}}}
\definmath{\qkt}{\tilde{\rm{q}}}
\definmath{\elt}{\tilde{\ell}}
\definmath{\hgt}{\tilde{\rm{H}}}
\definmath{\nut}{\tilde{\nu}}
\definmath{\dnt}{\tilde{\rm{d}}}
\definmath{\ftl}{\mr{\tilde{\rm{f}}}}
\definmath{\psb}{\bar{\psi}}
\definmath{\rtt}{\sqrt{2}}
\definmath{\mut}{\tilde{\mu}}
\newc{\mr}{\mathrm}
\newc{\bath}{\bar{\theta}}
\newc{\tht}{\theta}
\newc{\JC}{{\bf J}}
\newc{\lra}{\longrightarrow}
\newc{\eg}{{\it e.g.\  }}
\newc{\barr}{\begin{eqnarray}}
\newc{\earr}{\end{eqnarray}}
\newc{\me}{\mathcal{M}}
\definmath{\dbm}{\partial_\mu}
\definmath{\dbmu}{\stackrel{\leftrightarrow\  }{\partial^\mu}}
\definmath{\sgm}{\sigma_\mu}
\newc{\captionB}[2]{\caption[{#1}]{{\small {#2}}}}
\title{Extracting the Flavour Structure of a
         Baryon-Number R-parity Violating Coupling at the LHC}

\author{B.C. Allanach$^*$, A.J. Barr$^\dagger$, M.A. Parker$^\dagger$,
 P. Richardson$^{\dagger,\ddagger}$ and B.R. Webber$^\dagger$\\
$^*$Theory Division, CERN, 1211 Geneva 23, Switzerland.\\
$^\dagger$Cavendish Laboratory, University of Cambridge, Madingley Road,
        Cambridge, CB3\nolinebreak\  \nolinebreak{0HE,} UK.\\
$^\ddagger$DAMTP, Centre for Mathematical Sciences, Wilberforce Road, Cambridge,
        \mbox{CB3~0WA,~UK}.
}
\abstract{The detection of the flavour content of jets 
produced from R-parity violating
neutralino decays is investigated in the case where
one baryon-number violating coupling dominates.
Simulations are performed of the ATLAS experiment at the LHC 
for all couplings, other than $\lampp_{{\rm t}jk}$
since neutralino decays through these couplings are very suppressed.
Secondary vertex distributions and muons produced by 
heavy-quark \mbox{(b- and c-)} jets 
allow discrimination between LSP decay modes.
The dominant coupling can be identified at better than 3.5~$\sigma$ 
in almost all cases, with the only remaining ambiguity 
caused by the inability to distinguish strange from down quarks.
}
\preprint{Cavendish HEP-2001-07\\
        DAMTP-2001-48\\
        ATLAS-COM-PHYS-2001-010\\
        CERN-TH-2001-155}
\begin{document}
\section{Introduction}

In supersymmetry (SUSY) models where R-parity,
$\rp=(-1)^{3\rm{B}+\rm{L}+2\rm{S}}$, is violated (RPV models) the
lightest supersymmetric particle (LSP) can decay to Standard Model particles.
For the baryon-number violating RPV coupling, \lampp, the LSP,  
assumed to be the \ntlone, decays via 
a virtual squark into three quarks (or antiquarks).
In previous work \cite{Allanach:2001xz} we suggested a method by which
the masses of several of the SUSY particles can be measured in such
scenarios for a wide range of \lampp.

If R-parity violating couplings are observed, 
it seems probable that they come from an underlying theory which also
determines the fermion mass spectrum. This idea has been
explored~\cite{Ellis:1998rj, Chkareuli:1999at, Leontaris:1999wf, Ellis:2000js} 
in the context of gauged $U(1)_X$ models of flavour, in which a hierarchical 
pattern of RPV couplings is generated by the same symmetry from which the 
fermion mass spectrum is derived. 
Thus the determination of the flavour structure of an RPV coupling could be 
a vital clue in the resolution of the flavour problem.

In this paper we investigate the flavour content of the quark 
jets from \ntlone\  decays. The aim is to identify which $\lampp_{ijk}$ 
coupling is dominant in the neutralino decay.
The index $i$ is the generation number of an up-type quark, and  
$j$ and $k$ correspond to down-type quarks. The coupling is 
antisymmetric in $j$ and $k$, so there are nine possible non-zero elements.

Three of the couplings, $\lampp_{{\rm t}jk}$, produce very 
different signatures because they allow the \ntlone\ to decay via the top quark.
If the neutralino is lighter (heavier) than the top quark then the decay is 
propagator (phase-space) suppressed.
For the mSUGRA point we investigate (which is described in Section~\ref{sect:sim})
the branching ratio through the top coupling is a factor of $4.2 \times 10^6$ 
smaller than through a non-top coupling of the same size.

For a top mode to dominate the decay, a \lampp\ coupling to top would need 
to be a factor of at least 2000 greater than any of the non-top couplings.
In that case the lightest neutralino would
typically be long-lived (Figure~\ref{fig:life}) and
for $\lampp_{{\rm t}jk}\lsim 10^{-2}$ nearly all \ntlone\ s decay
outside the detector. Determination of these couplings could then only 
be achieved by searching for the rare decay of $\ntlone$s in the active volume, 
while sparticle mass measurements could be made in the same way as for 
the \rp\ conserving case\cite{atlastdr}.

\EPSFIGURE[t]{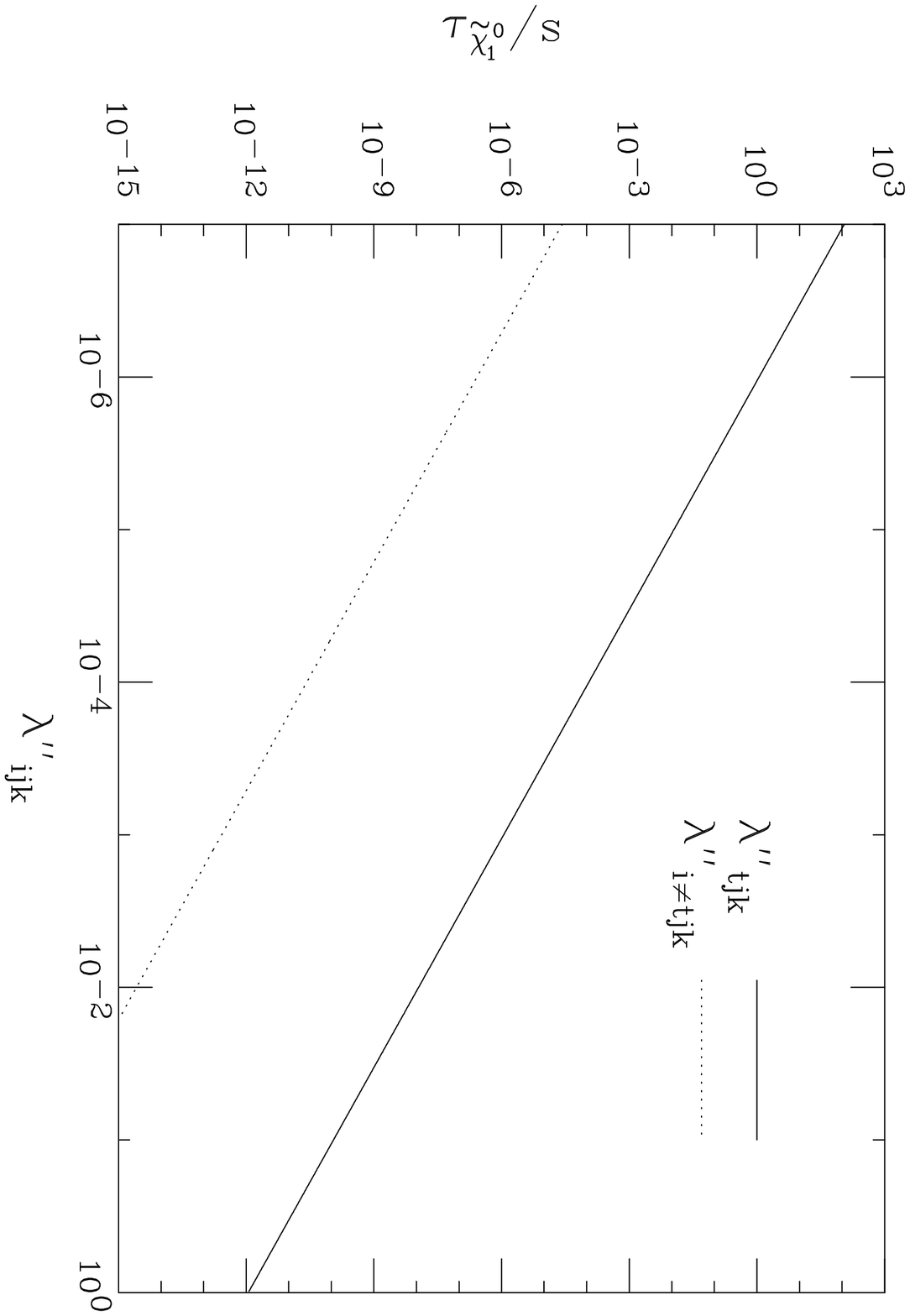, height=12cm, angle=+90}{
The lifetime of the \ntlone\ 
at the mSUGRA point described in the text,
with one of $\lampp_{ijk}$ non-zero.
The solid line shows the lifetime for a dominant
top quark coupling, which leads to the decay 
$\ntlone\to\rm{qq't^*}\to\rm{qq'bW^*}\to\rm{qq'bff'}$, 
where q and $\rm q'$ are quarks other than top and f and $\rm f'$ 
are Standard Model fermions.
The dashed line shows the lifetime for couplings to quarks other than
the top.
The calculation approximates the final state particles as massless, so
that the lines for $\lampp_{{\rm t}jk}$ are the same for all $j,k$, 
as are those for $\lampp_{ijk}$ for all~$i\ne {\rm t},j,k$.
\label{fig:life}
}

We investigate the other six couplings --
$\lampp_{\rm{uds}}$, $\lampp_{\rm{udb}}$, $\lampp_{\rm{usb}}$, 
$\lampp_{\rm{cds}}$, $\lampp_{\rm{cdb}}$ and $\lampp_{\rm{csb}}$ --
assuming that one is significantly larger than the others.
Since the branching ratio is proportional to the square of the coupling,
it is sufficient for this analysis that all the subdominant couplings 
are 4.5 times smaller than the dominant coupling. Then just 5\% 
of all $\ntlone$s will decay though the subdominant channel, and 90\% 
of events, each with two $\ntlone$s, will contain only dominant-coupling decays.

\section{Event Simulation and Selection}
\label{sect:sim}
HERWIG~6.2\cite{Marchesini:1991ch,Corcella:2000bw,
                Dreiner:1999qz,Richardson:2000nt}
is used as the event generator, and the official ATLAS simulation
program ATLFAST~2.50 \cite{Richter:1998at}, is used to simulate the
performance of the ATLAS detector. We use as a test case the mSUGRA point 
\mbox{$m_0=100$}~GeV, \mbox{$m_{1/2}=300$}~GeV, \mbox{$A_0=300$}~GeV,
\mbox{$\tan\beta=10$} and 
\mbox{$\rm{sgn}\,\mu\ +$}, with one of \mbox{$\lampp_{ijk}=5\times10^{-3}$}.
This is in the middle of the range of couplings $10^{-5}\lsim\lampp\lsim 1$,
for which we can expect double \ntlone\ decay within the detector, 
and is approaching indirect limits set from nuclear decays for
$\lampp_{\rm{uds}}$ and $\lampp_{\rm{udb}}$ \cite{Sher:1994sp, Goity:1995dq, Allanach:1999ic}.
A selection of the sparticle masses for this point 
is shown in Table~\ref{tab:mass}.
The applicability of our analysis to SUGRA models with other parameters 
is discussed in our previous work~\cite{Allanach:2001xz}.
The method of the event selection, which is also detailed in that paper, 
is summarised below.

\renewcommand{\arraystretch}{1.2}
\TABULAR[b]{|l|l|l|l|l|l|l|l|l|}{
\hline
\ntlone & \ntltwo & \glt & $\upt_R$ & $\upt_L$ & 
$\dnt_R$ & $\dnt_L$ & $\elt_R$ & $\elt_L$ \\ 
\hline
116.7 &         211.9 &         706.3 &         611.7 & 632.6 & 610.6 & 637.5 & 155.3 & 230.5 \\        
\hline
}{
\label{tab:mass}
Masses of selected particles (in GeV) for the model investigated.
These were calculated from the mSUGRA parameters using
ISASUGRA from the package ISAJET~7.51\cite{Baer:1999sp}.
}
\renewcommand{\arraystretch}{1.0}

Signal events will contain three jets from each of the two $\ntlone$s. Since
the SUSY cross section is dominated by squark and gluino production, 
there are usually at least two other jets from the \rp\ conserving decays of 
squarks and gluinos.
Requiring the presence of leptons, which can originate in the decay chain
$\ntltwo \rightarrow \sslr^\mp\ell^\pm \rightarrow \ntlone\ell^\mp\ell^\pm$,
helps to decrease the Standard Model (SM) background.
Events are selected with between eight and ten jets, and with 
two opposite-sign, same family isolated leptons (e or $\mu$).

Cuts on event-shape variables are applied, reducing the 
SM background to less than 10\% of the SUSY signal, 
and kinematic cuts are applied to 
preferentially select jets from neutralino decays. All possible 
three-jet combinations are inspected and their invariant masses calculated.
Since we expect the event to contain two LSP decays
we select two of these three-jet combinations with
reconstructed masses within 20~GeV of each other as \ntlone\ candidates.
To reconstruct the \ntltwo\ we pick the \ntlone\ candidate 
closest in $\eta-\phi$ to either of the leptons 
from the lepton pair, and find the total invariant
mass of the three jets plus those two leptons.

A 2-dimensional Gaussian is fitted to the \ntlone\ and \ntltwo\ masses, 
and combinations are selected within $2 \times \sigma$ of that peak.
This selection removes much of the combinatoric background, so that the jet 
multiplicity cut can be relaxed to $8\le N_{\rm{jet}} \le 12$.
We now have a rather clean sample of events in which we
can identify six jets with the two neutralino decays.
All $N_{\rm comb}$ jet combinations which pass the above cuts
for any event are accepted and are given weight $1/N_{\rm comb}$.

\section{Vertex Tagging}
\label{sec:ver}
The lifetimes of hadrons containing \mbox{b- and c-quarks} give rise to 
displaced vertices
which can be reconstructed from charged tracks in the inner detector.
In general the lifetimes of charmed hadrons are shorter than 
for hadrons containing a b-quark, 
for example $c\tau_{D^0} = 124 ~\mu \rm{m}$ compared to 
$c\tau_{B^0} = 468 ~\mu \rm{m}$. 
This allows statistical separation of c-quark jets from b-quark jets, 
and of c- and b-jets from light quark (u, d, s) and gluon jets.
Strange hadrons have longer lifetimes with $c\tau$ of the order of 
tens of centimeters.
Since they decay to a small number of particles it is difficult to 
reconstruct a secondary vertex, and so strange hadrons
cannot be tagged within jets in the LHC environment.

\EPSFIGURE[t]{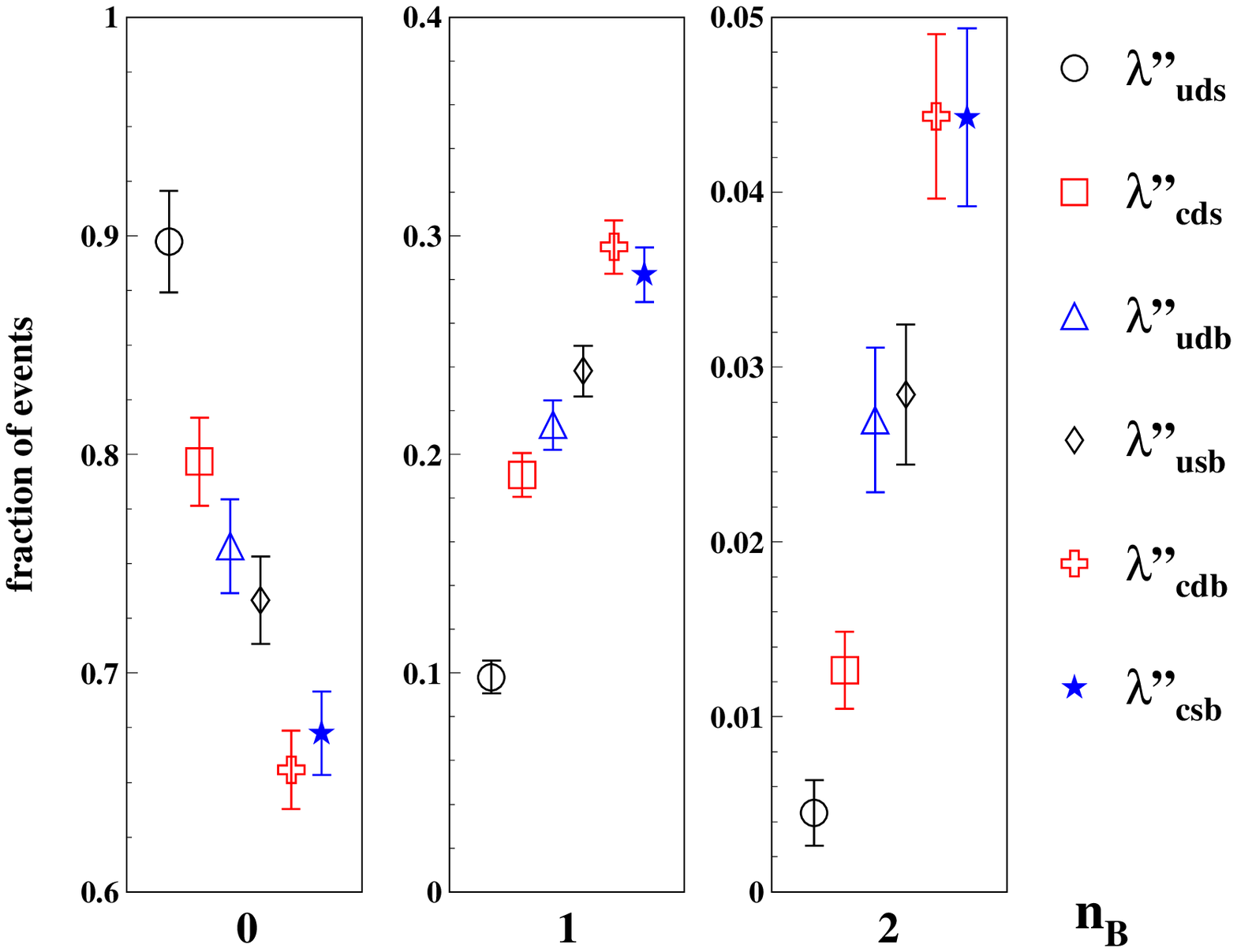, height=11.5cm}{
The fraction of signal events with 0, 1, and 2 vertex tags, for different
types of RPV coupling \lampp.
The statistical errors are those expected for an integrated luminosity of $30~\rm{fb}^{-1}$.
Points are horizontally displaced to allow them to be distinguished by eye.
\label{fig:btag}
}

\TABULAR[t]{|c|c|c|}{
\hline
{b}-tagging efficiency &
\multicolumn{2}{c}{Rejection factor} \vline \\
\cline{2-3}
$\epsilon_b$ & u, d, s and g jets $(r_j)$ & c-jets $(r_c)$ \\
\hline
0.33 & 1400 & 22.9 \\
\hline
}{
\label{tab:btag}
b-tagging efficiencies and mis-tagging rejection factors from a full simulation
of the ATLAS inner detector\cite{atlastdr}.
$1/r_c$ is the probability of tagging a c-jet, while 
$1/r_j$ is the probability of tagging a u, d, s or gluon jet, averaged
over $p_T$, $\eta$ and $\phi$. The typical reconstructed jet $p_T$ 
scale is 50~GeV.
}

The vertex tagging performance of the ATLAS Inner Detector was
simulated in \cite{atlastdr}.
In that study a likelihood ratio method was applied to
the transverse impact parameter $(d_0)$ of the tracks within the jet cone.
The resultant b-tagging efficiency and the rejection rates for 
c-~and other jets were parameterised as functions of the
transverse momentum and pseudorapidity of the reconstructed jet and were
implemented in ATLFAST.
The rejection factors, averaged over all directions and transverse momenta,
are summarised in Table~\ref{tab:btag}.
In order to select a higher purity sample, we
tolerate a rather low b-tagging efficiency (33\%).

The proportions of events with \mbox{0, 1 and 2} tagged jets are 
shown in Figure~\ref{fig:btag}, 
for the six different \lampp\  couplings. As we expect, the vertex tagging rate is
greatest when the \ntlone\  decay products include both \mbox{c and b} quarks, 
and smallest for couplings which produce light-quark daughters.
As we cannot distinguish d- from s-jets,
there is no discrimination between $\lampp_{i{\rm db}}$ and $\lampp_{i{\rm sb}}$
(for $i$=u,c).

In these simulations, with $\lampp=5\times10^{-3}$ the \ntlone\  lifetime 
of $10^{-14}$ seconds corresponds to a typical decay length $c\tau$ 
of $3~\mu {\rm m}$. This is short in comparison with typical 
b- and c-jet vertex displacements so it should not severely 
affect vertex reconstruction.
However, the lifetime is inversely proportional to the square of the 
coupling strength. When $c\tau$ becomes bigger than about 3~cm 
(for $\lampp \lsim 5\times 10^{-5}$) special vertex reconstruction 
would be required.
Vertex tagging becomes almost impossible when neutralinos 
travel more than about 30~cm in 
the transverse direction since their daughters will produce hits 
only in the outermost silicon layers.

\section{Flavour Discrimination from Muons}

\EPSFIGURE{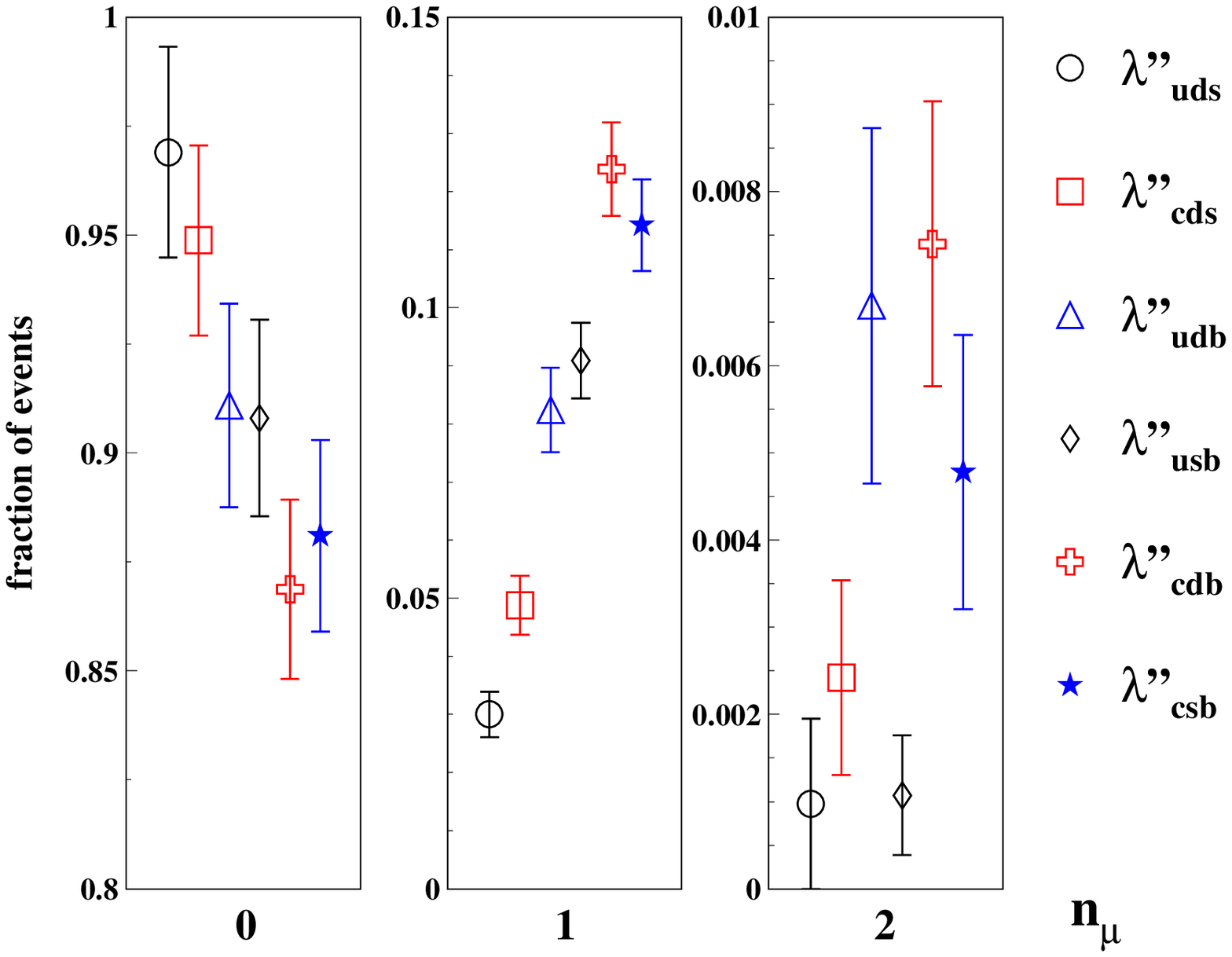, height=11.5cm}{
The fraction of signal events with \mbox{0, 1 and 2} muons that lie within 
$\sqrt{\Delta\eta^2-\Delta\phi^2} \le 0.4$ of the signal jets, 
for different types of RPV coupling, $\lampp$.
The statistical errors are those expected for an integrated luminosity of $30~\rm{fb}^{-1}$.
Points are horizontally displaced to allow them to be distinguished by eye.
\label{fig:mux}
}

The weak decay of hadrons containing heavy quarks can produce muons, 
which will generally lie within the associated quark jet.
The muons will pass though the calorimeter and so can be measured by 
the muon detector even if they lie inside the jet cone. 
The frequency with which these ``non-isolated'' muons occur in signal
events can be used as an additional discriminator between quark flavours.

The number of non-isolated muons per event in the six signal jets is plotted in 
Figure~\ref{fig:mux}. Since bottom mesons produce more muons 
[$B(B^0\to \mu^+ \ \nu_{\mu} \ {\rm X}) = 10\%$] 
than charmed mesons [$B(D^0\to \mu^+ \ \nu_{\mu} \ {\rm X}) = 7\%$]
\cite{Caso:1998tx},
we can use the number of muons in signal jets to
statistically separate \mbox{b- from c-jets}.

\section{Statistical Significance}

\TABULAR[t]{|c|c|c|c|c|c|c|}{
\hline
\multicolumn{2}{|c}{Distinguishing} \vline & 
\multicolumn{2}{c}{Vertexing} \vline &
\multicolumn{2}{c}{Muons} \vline & Combined \\
\cline{3-6}
\multicolumn{2}{|c}{$\lampp_{ijk}$ from $\lampp_{lmn}$} \vline & 
$\chi^2$/d.f.  & P/\% & $\chi^2$/d.f. & P/\% & discrimination/$\sigma$\\
\hline
uds & udb & 59.1/1 & - & 28.7/1 & - & 9.4\\
 & usb & 73.0/1 & - & 31.7/1 & - & 10.2\\
 & cds & 30.5/1 & - & 4.0/1 & 4 & 5.9\\
 & cdb & 106.9/1 & - & 47.2/1 & - & 12.4\\
 & csb & 113.4/1 & - & 49.2/1 & - & 12.8\\
\hline
udb & usb & 1.6/2 & 44 & 0.4/1 & 54 & 1.4\\
 & cds & 10.3/2 & 1 & 13.0/1 & - & 4.8\\
 & cdb & 18.3/2 & - & 6.8/2 & 3 & 5.0\\
 & csb & 16.3/2 & - & 5.1/2 & 8 & 4.6\\
\hline
usb & cds & 17.5/2 & - & 17.2/1 & - & 5.9\\
 & cdb & 12.1/2 & - & 5.1/1 & 2 & 4.2\\
 & csb & 9.9/2 & 1 & 3.1/1 & 8 & 3.6\\
\hline
cds & cdb & 56.1/2 & - & 37.4/1 & - & 9.7\\
 & csb & 55.8/2 & - & 35.3/1 & - & 9.5\\
\hline
cdb & csb & 0.6/2 & 72 & 1.3/2 & 51 & 1.4\\
\hline
}{
\label{tab:chi}
Chi-squared function, and number of degrees of freedom (d.f.)
for the difference in distributions for pairs of RPV couplings 
(\lampp) for an integrated luminosity of $30~{\rm fb}^{-1}$. 
The contributions from vertexing and muon counting
are shown separately. The probability (P) in the tail of the chi-squared
distribution is given when ${\rm P}\ge 1\%$. The number of degrees of freedom is
one less than the number of histogram bins in which both couplings 
contain at least five events.
}

The confidence with which we can identify the dominant coupling
was explored for all pairs of couplings for both the vertex tagging 
and muon rates (Figures~\ref{fig:btag} and \ref{fig:mux} respectively). 
The variable,
\barr
\chi^2=\sum_i\frac{(x_i-y_i)^2}{\sigma^2_{x}+\sigma^2_{y}}
\nonumber
\earr
was calculated, where
$x_i$ and $y_i$ are the fractions of events with $i$ muons 
(or vertex tags) for the test pair of couplings
and the $\sigma^2$s are their variances assuming Poisson-distributed 
numbers of events. Bins where one or other distribution contained
fewer than five events were excluded.

The calculated $\chi^2$ values for both the muon and vertex-tagging plots
are shown in Table~\ref{tab:chi}. 
If we use only the muon information then all
couplings are distinguishable at 90\% except for the ambiguity 
between \mbox{d- and s-jets}. In general the vertexing rate gives 
better discrimination, but would be difficult for a long-lived 
\ntlone\ for the reasons discussed in Section \ref{sec:ver}. 
Combining the results from both analyses gives separation at better 
than 3.5~$\sigma$ in all cases, again with the exception 
that it is not possible to distinguish down from strange quarks.

If the branching ratios of the neutralino through two or more 
couplings were comparable, then as the information in the
distributions is rather degenerate, definitive identification of
the couplings becomes difficult. 
However the method could then be used to constrain the possible
values of the couplings.

\section{Conclusions}
If baryon-number violating couplings to quarks other than top 
are $\gsim 5\times10^{-5}$ then $\ntlone$s will typically decay within
the tracking volume of an LHC detector.
For these couplings, our simulations show that 
displaced secondary vertices and muons from heavy quark decays allow 
statistical separation of \mbox{b- from c-quark} jets, 
and \mbox{b- and c-jets} from light-quark jets.
A dominant RPV coupling can then be identified at better than 3.5~$\sigma$,
except for an ambiguity caused by the inability to 
distinguish strange from down quarks. 

We have demonstrated that if the MSSM breaks baryon number, not only 
can the LHC detect and measure the masses of 
sparticles\cite{Allanach:2001xz}, but it can even distinguish the 
flavour structure of the RPV coupling.

\acknowledgments
We would like to thank F. Paige for suggesting this extension of our          
previous work. 
We thank H. Dreiner, members of the Cambridge SUSY
working group, and the ATLAS collaboration for helpful discussions.
We have made use of the physics analysis framework and tools which are
the result of collaboration-wide efforts. This work was partly funded by PPARC.

\bibliography{flavour}
\end{document}